# Analytical Study of Hybrid Surface Phonon-Plasmon-Polaritons ($HSP^3$) in Symmetric Nonlinear-Graphene-hBN Heterostructures

**Mohammad Bagher Heydari [1,*], Majid Karimipour [2], Morteza Mohammadi Shirkolaei [3]**

[1,*] School of Electrical Engineering, Iran University of Science and Technology (IUST), Tehran, Iran
[2] Department of Electrical Engineering, Arak University of Technology, Arak, Iran
[3] Department of Electrical Engineering, Shahid Sattari Aeronautical University of Science and Technology, Tehran, Iran

[*]Corresponding author: mo_heydari@alumni.iust.ac.ir

**Abstract:** The hybridization of hyperbolic Polaritonics with THz Plasmonics has attracted immense attention due to its fascinating applications in this region. However, to effectively enhance the performance of these coupled modes, one way is the usage of smart materials in the heterostructures. Here, we introduce a symmetric graphene-based structure containing hexagonal Boron Nitride (hBN) layers, where two nonlinear layers have been utilized as claddings to improve the confinement of Hybrid Surface Phonon-Plasmon-Polaritons ($HSP^3$). A new analytical model is derived by using Maxwell's equations and applying the boundary conditions. A set of nonlinear equations are solved numerically and the obtained results are reported. Harnessing a nonlinear medium together with hybrid graphene-hBN layers allows one to tune the propagation properties of the structure by changing the chemical potential, the relaxation time, and the nonlinear factor. The study has been done for two frequency regions: the upper Reststrahlen and the lower Reststrahlen bands. The numerical results show a long propagation length ($L_{prop}$=300 μm) and a large Figure of Merit (FOM=98) for $HSP^3$ propagating in the upper Reststrahlen band at the frequency of 45 THz. The proposed structure and its analytical model can open new insight into the design of tunable devices in mid-infrared plasmonics.

**Key-words:** Graphene, hBN, nonlinear medium, plasmon, phonon, analytical

## 1. Introduction

Polaritons are hybrid modes resulting from the coupling among the charge dipoles and photons. The common kinds of these modes are Surface Plasmon Polaritons (SPPs) [1, 2] and Phonon Polaritons [3-5]. SPPs are formed by strong coupling of the light and free electrons, which have many interesting applications such as couplers [6-8], filters [9-11], resonators [12-14], and circulators [15-18], waveguide [19-25], sensors [26-31], metamaterials [32], and imaging [33, 34]. Among these devices, graphene-based waveguides play a remarkable role in plasmonics and have various platforms such as planar [21, 35-45], cylindrical [46-52], and elliptical structures [22, 53-55]. Phonon polaritons have a different origin compared to SPPs; they are excited by the lattice vibrations in polar crystals like hBN [56].

In the mid-infrared region, graphene is one of the promising materials to design innovative devices, due to its tunable conductivity [57, 58]. Graphene can support highly-confined SPPs with long propagation lengths, compared to metal-based devices [51, 59]. An effective method to enhance the performance of graphene-based components is by integrating graphene with new smart materials such as chiral materials [23, 60-66], and non-linear materials [67-77]. The hybridization of graphene with a nonlinear medium enhances the quality of propagating SPPs [59, 67, 69, 78-80]. For instance, a graphene-based Kerr-type structure is reported in [78], in which SPPs have high FOM. In [67], a nonlinear magnetic cladding in a graphene-based waveguide is utilized to adjust the propagating features via the nonlinear coefficient. The tunability of hybrid plasmons by varying the graphene parameters of nonlinear-graphene structure is shown in [81]. In [80], the authors have derived a dispersion relation for a nonlinear medium incorporating a graphene sheet and have reported an enhancement of field localization of hybrid plasmons by changing the nonlinear permittivity.

The hBN is a popular polar crystal, supporting phonon polaritons in the mid-infrared region due to its natural hyperbolicity [3]. It has been demonstrated that these phonon polaritons in hBN have many fascinating applications such as data storage [82], microscopy [83], and negative refraction [84]. Both SPPs on graphene and phonon polaritons on hBN are excited in the mid-infrared frequencies; so the hybridization of these materials can be interesting [56, 85, 86]. This hybridization introduces a new class of polaritons called "Hybrid Surface Phonon-Plasmon-Polaritons ($HSP^3$)" [87], with novel applications such as enhancing mobility [88], and phonon-induced transparency [89], negative refractive [90], planar focusing [91], and topological transition [92]. The basic research on $HSP^3$ in graphene-hBN heterostructures is reported by Anshuman Kumar et al. [56] and they have studied theoretically the propagating plasmon-phonon modes in typical air-graphene-hBN-air (AGHA) and air-graphene-hBN-graphene-air (AGHGA) systems. In [87], mid-infrared measurements are performed on nano-resonators to examine the excitation of plasmon coupling to phonon modes. The authors in [93] introduced a new kind of $HSP^3$ excited by an acoustic wave on a piezoelectric substrate. In [94], a new THz metamaterial constructed of multilayer graphene-hBN structures is investigated for TE/TM polarizations. A similar study on the propagating features of $HSP^3$ in graphene-hBN metamaterials is demonstrated by Haoyuan Song et al. [95], where an epsilon-near-zero property is reported.

In this paper, to enhance the localization of $HSP^3$ in graphene- hBN heterostructure and achieve higher values of FOM, we introduce a new platform based on integrating graphene-hBN layers with nonlinear media. This novel heterostructure can effectively change and control the supporting $HSP^3$ and also gives us a better performance. To the best of our knowledge, no published work is reported for the study of $HSP^3$ in the nonlinear area. The paper is organized as follows. In section 2, the proposed structure is introduced first and then its analytical model is presented by embarking on Maxwell's equations and applying the boundary conditions. Then, the numerical results solved by MATLAB software are investigated and discussed in section 3. Finally, section 4 concludes the paper.

## 2. The proposed structure and its analytical model

Fig. 1 illustrates the configuration of the proposed structure, where two nonlinear layers on both sides of the whole structure have been utilized as claddings. As seen in this figure, graphene sheets have been located at the border of hexagonal boron nitride (hBN) and nonlinear layers. Graphene conductivity can be modeled by popular relation in the literature, called Kubo's formula [96]:

$$\sigma(\omega)=\frac{-je^2\omega}{\pi}\left[\frac{2}{\omega^2}\int_0^\infty \varepsilon d\varepsilon\left(\frac{df_0(\varepsilon)}{d\varepsilon}\right)-\int_0^\infty d\varepsilon\frac{f_0(-\varepsilon)-f_0(\varepsilon)}{\omega^2-4\varepsilon^2}\right] \tag{1}$$

In (1), $f_0(\varepsilon)$ is the Fermi-Dirac distribution, ω is radian frequency, and e is the electron charge in this relation. In the above relation, the first term is related to intra-band transition while the remaining terms express the inter-band contribution.

In fig. 1, the nonlinear layers have the following permittivity [97]:

$$\varepsilon_{NL}=\varepsilon_L+\alpha\left|\boldsymbol{E}\right|^2 \tag{2}$$

The above relation can be approximately rewritten as follows:

$$\varepsilon_{NL}\approx\varepsilon_L+\alpha\eta^2\left|\boldsymbol{H}\right|^2\approx\varepsilon_L+\alpha'\left|\boldsymbol{H}\right|^2 \tag{3}$$

Consider TM mode propagating in the x-direction ( $e^{j\beta x}$) inside the structure with the electromagnetic components of $E_x, E_z, H_y$. Thus, the relation (3) can be expressed as:

$$\varepsilon_{NL}=\varepsilon_L+\alpha'\left|H_y\right|^2 \tag{4}$$

Where

$$\alpha'=\frac{\alpha}{\varepsilon_0\varepsilon_L c^2} \tag{5}$$

is a nonlinear coefficient defined in (4).

To present an analytical model for the proposed structure shown in Fig. 1, Maxwell's equations should be written inside each medium and then we should apply boundary conditions for the electromagnetic fields. To do this, first, consider Maxwell's equations inside the nonlinear media (suppose $e^{-j\omega t}$):

$$\nabla\times\boldsymbol{E} = j\omega\mu_0\boldsymbol{H} \tag{6}$$

$$\nabla\times\boldsymbol{H} = -j\omega\varepsilon_0\varepsilon_{NL}\boldsymbol{E} \tag{7}$$

By substituting the field components of TM mode (i.e. $E_x, E_z, H_y$) in (6)-(7), one can obtain the following nonlinear equation:

$$\frac{d^2H_y}{dz^2} - \gamma_{NL}^2 H_y + k_0^2\alpha' H_y^3 = 0 \tag{8}$$

Where $k_0$ is the free-space wavenumber and,

$$\gamma_{NL}^2 = \beta^2 - k_0^2\varepsilon_L \tag{9}$$

is the wave constant. In this medium, the transverse component of electric fields can be expressed as follows:

$$E_x = \frac{1}{j\omega\varepsilon_0\varepsilon_{NL}}\frac{\partial H_y}{\partial z} \tag{10}$$

$$E_z = \frac{-\beta}{\omega\varepsilon_0\varepsilon_{NL}}H_y \tag{11}$$

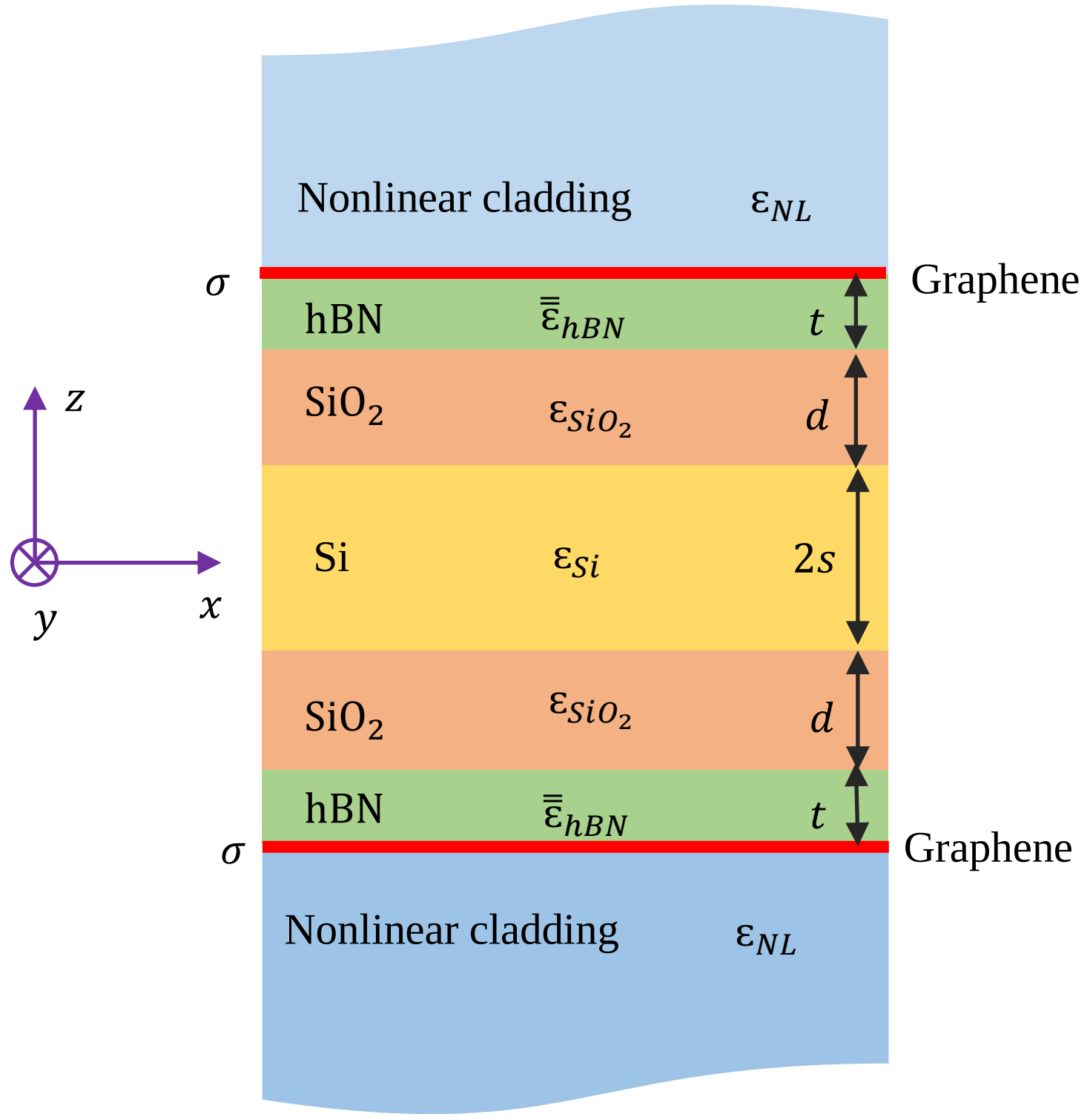

**Fig. 1.** The schematic of the proposed structure: a symmetric structure, where two nonlinear layers at both sides of the whole waveguide have been utilized as cladding layers. The graphene sheet has been located at the border of nonlinear material and hexagonal boron nitride (hBN) crystal.

The first integral of equation (8) will give us:

$$\left[\frac{dH_y}{dz}\right]^2 - \gamma_{NL}^2 H_y^2 + \frac{1}{2}k_0^2\alpha' H_y^4 = S \tag{12}$$

In (12), $S$ is an integration constant. By supposing $S \geq 0$ (for the case of $S \leq 0$, the expressions are similar), we will obtain:

$$\frac{dH_y}{dz} = \pm\sqrt{S + \gamma_{NL}^2 H_y^2 - \frac{1}{2}k_0^2\alpha' H_y^4} \tag{13}$$

In general, the solution of (13) is written by Jacobi elliptic function (denoted by “cn”) and its modulus ($r$) [98]:

$$H_y(z) = A_{NL}\, cn\left(q\left[(z - z_0) + z_{oc}\right], r\right) \tag{14}$$

In (14), the following parameters have been utilized:

$$q = \sqrt[4]{\gamma_{NL}^4 + 2S\,k_0^2\alpha'} \tag{15}$$

$$r = \frac{q^2 + \gamma_{NL}^2}{2q^2} \tag{16}$$

$$A_{NL} = \frac{1}{k_0}\sqrt{\frac{q^2 + \gamma_{NL}^2}{\alpha'}} \tag{17}$$

As a special case of (14), for $S = 0, r = 1$, we have:

$$H_y(z) = \frac{\gamma_{NL}}{k_0}\sqrt{\frac{2}{\alpha'}}\, sech\left(\gamma_{NL}(z - z_0)\right) = A_{NL}\, sech\left(\gamma_{NL}(z - z_0)\right) \tag{18}$$

Now, the analytical relations inside the nonlinear media are completed. In what follows, we will consider Maxwell’s equations inside the hBN. It should be noted that hBN is a popular anisotropic crystal with the following permittivity tensor [56]:

$$\varepsilon_m(\omega) = \varepsilon_{\infty,m} + \varepsilon_{\infty,m}.\frac{(\omega_{LO,m})^2 - (\omega_{TO,m})^2}{(\omega_{TO,m})^2 - \omega^2 - j\omega\Gamma_m} \tag{19}$$

In (19), $m = \parallel or \perp$ is related to the transverse and z-axis, respectively. Moreover, $\omega_{LO}, \omega_{TO}$ show the LO and TO phonon frequencies, respectively, in which each frequency has two values in the upper and lower Reststrahlen bands: $\omega_{LO,\perp} = 24.9\,THz, \omega_{TO,\perp} = 23.4\,THz, \omega_{LO,\parallel} = 48.3\,THz,\ \omega_{TO,\parallel} = 41.1\,THz$. In (19), $\Gamma_m$ is a damping factor ($\Gamma_\perp = 0.15\,THz, \Gamma_\parallel = 0.12\,THz$) and $\varepsilon_m$ is related to the high-frequency permittivity ($\varepsilon_{\infty,\perp} = 4.87, \varepsilon_{\infty,\parallel} = 2.95$) [56].

The first Maxwell’s equation inside the hBN is similar to that inside the nonlinear media (see relation (6)), while the second equation (i.e. the equation (7)) can be rewritten as [99]:

$$\nabla \times \boldsymbol{H} = -j\omega\varepsilon_0 \overset{=}{\varepsilon}.\boldsymbol{E} \tag{20}$$

Again, by considering TM propagating mode inside hBN (i.e. $E_x, E_z, H_y$), we finally obtain the following wave equation and its wave constant,

$$\frac{d^2 H_y}{dz^2} - \gamma_{hBN}^2 H_y = 0 \tag{21}$$

$$\gamma_{hBN}^2 = \beta^2 \frac{\varepsilon_\perp}{\varepsilon_\parallel} - k_0^2\varepsilon_\perp \tag{22}$$

Similar to the relations (10)-(11), the transverse components of electric fields are derived for the nonlinear medium:

$$E_x = \frac{1}{j\omega\varepsilon_0\varepsilon_\perp}\frac{\partial H_y}{\partial z} \tag{23}$$

$$E_z = \frac{-\beta}{\omega\varepsilon_0\varepsilon_{||}} H_y \tag{24}$$

The electromagnetic analysis for the field components of hBN is completed now. It should be emphasized that the above relations ((21)-(24)) are utilized for dielectric media (Si and $SiO_2$ layers) by substituting $\varepsilon_\perp = \varepsilon_{||} = \varepsilon_r$. Now, for the proposed waveguide shown in fig. 1, the magnetic component inside each layer can be written as follows:

$$H_y(z) = \begin{cases} A_{NL}\ \mathrm{sech}\left(\gamma_{NL}\left(z+z_1\right)\right) & z > s+d+t \\ B e^{-j\gamma_{hBN}z} + C e^{+j\gamma_{hBN}z} & s+d < z < s+d+t \\ D e^{-j\gamma_{SiO_2}z} + E e^{+j\gamma_{SiO_2}z} & s < z < s+d \\ F e^{-j\gamma_{Si}z} + G e^{+j\gamma_{Si}z} & -s < z < s \\ H e^{-j\gamma_{SiO_2}z} + I e^{+j\gamma_{SiO_2}z} & -(s+d) < z < -s \\ J e^{-j\gamma_{hBN}z} + K e^{+j\gamma_{hBN}z} & -(s+d+t) < z < -(s+d) \\ A_{NL}\ \mathrm{sech}\left(\gamma_{NL}\left(z+z_2\right)\right) & z < -(s+d+t) \end{cases} \tag{25}$$

In (25), $B, C, D, E, F, G, H, I, J, K$ are unknown coefficients and $A_{NL}$ is given in (18). These unknown coefficients should be obtained by applying the boundary conditions. From (10) and (23), the $E_x$ component inside each layer is derived:

$$E_x(z) = \frac{-1}{\omega\varepsilon_0}\begin{cases} \left(-jA_{NL}\frac{\gamma_{NL}}{\varepsilon_{NL}}\right)\mathrm{sech}\left(\gamma_{NL}\left(z+z_1\right)\right)\times\tanh\left(\gamma_{NL}\left(z+z_1\right)\right) & z > s+d+t \\ \frac{\gamma_{hBN}}{\varepsilon_{hBN}}\left(Be^{-j\gamma_{hBN}z} - Ce^{+j\gamma_{hBN}z}\right) & s+d < z < s+d+t \\ \frac{\gamma_{SiO_2}}{\varepsilon_{SiO_2}}\left(De^{-j\gamma_{SiO_2}z} - Ee^{+j\gamma_{SiO_2}z}\right) & s < z < s+d \\ \frac{\gamma_{Si}}{\varepsilon_{Si}}\left(Fe^{-j\gamma_{Si}z} - Ge^{+j\gamma_{Si}z}\right) & -s < z < s \\ \frac{\gamma_{SiO_2}}{\varepsilon_{SiO_2}}\left(He^{-j\gamma_{SiO_2}z} - Ie^{+j\gamma_{SiO_2}z}\right) & -(s+d) < z < -s \\ \frac{\gamma_{hBN}}{\varepsilon_{hBN}}\left(Je^{-j\gamma_{hBN}z} - Ke^{+j\gamma_{hBN}z}\right) & -(s+d+t) < z < -(s+d) \\ \left(-jA_{NL}\frac{\gamma_{NL}}{\varepsilon_{NL}}\right)\mathrm{sech}\left(\gamma_{NL}\left(z+z_2\right)\right)\times\tanh\left(\gamma_{NL}\left(z+z_2\right)\right) & z < -(s+d+t) \end{cases} \tag{26}$$

By applying the boundary conditions given in the following relations

$$E_{x,2} = E_{x,1} = E_x \tag{27}$$

$$H_{y,2}-H_{y,1}=\begin{cases}\sigma E_x & if\ z=|s+d+t| \\ 0 & otherwise\end{cases} \tag{28}$$

And doing some mathematical procedures, we achieve a system of nonlinear equations:

$$\begin{cases} a_1\, f_{NL,z_1}+a_2B+a_3C=0 \\ a_4B+a_5C+a_6D+a_7E=0 \\ a_8D+a_9E+a_{10}F+a_{11}G=0 \\ a_{11}F+a_{10}G+a_9H+a_8I=0 \\ a_7H+a_6I+a_5J+a_4K=0 \\ a_3J+a_2K-a_1\, f_{NL,z_2}=0 \\ a_{12}B+a_{13}C-g_{NL,z_1}=0 \\ a_{14}B+a_{15}C+a_{16}D+a_{17}E=0 \\ a_{18}D+a_{19}E+a_{20}F+a_{21}G=0 \\ a_{21}F+a_{20}G+a_{19}H+a_{18}I=0 \\ a_{17}H+a_{16}I+a_{15}J+a_{14}K=0 \\ a_{13}J+a_{12}K-g_{NL,z_2}=0 \end{cases} \tag{29}$$

With the following coefficients,

$$a_1=-jA_{NL}\frac{\gamma_{NL}}{\varepsilon_{NL}},\ a_2=-\frac{\gamma_{hBN}}{\varepsilon_{hBN}}e^{-j\gamma_{hBN}(s+d+t)},\ a_3=\frac{\gamma_{hBN}}{\varepsilon_{hBN}}e^{+j\gamma_{hBN}(s+d+t)},\ a_4=\frac{\gamma_{hBN}}{\varepsilon_{hBN}}e^{-j\gamma_{hBN}(s+d)}$$
$$a_5=-\frac{\gamma_{hBN}}{\varepsilon_{hBN}}e^{+j\gamma_{hBN}(s+d)},a_6=-\frac{\gamma_{SiO_2}}{\varepsilon_{SiO_2}}e^{-j\gamma_{SiO_2}(s+d)},a_7=\frac{\gamma_{SiO_2}}{\varepsilon_{SiO_2}}e^{+j\gamma_{SiO_2}(s+d)},a_8=\frac{\gamma_{SiO_2}}{\varepsilon_{SiO_2}}e^{-j\gamma_{SiO_2}(s)}$$
$$a_9=-\frac{\gamma_{SiO_2}}{\varepsilon_{SiO_2}}e^{+j\gamma_{SiO_2}(s)},a_{10}=-\frac{\gamma_{Si}}{\varepsilon_{Si}}e^{-j\gamma_{Si}(s)},a_{11}=\frac{\gamma_{Si}}{\varepsilon_{Si}}e^{+j\gamma_{Si}(s)},a_{12}=\frac{\left(-\frac{\sigma\gamma_{hBN}}{\varepsilon_{hBN}}+1\right)}{A_{NL}}e^{-j\gamma_{hBN}(s+d+t)} \tag{30}$$
$$a_{13}=\frac{\left(+\frac{\sigma\gamma_{hBN}}{\varepsilon_{hBN}}+1\right)}{A_{NL}}e^{+j\gamma_{hBN}(s+d+t)},a_{14}=e^{-j\gamma_{hBN}(s+d)},a_{15}=e^{+j\gamma_{hBN}(s+d)},a_{16}=-e^{-j\gamma_{SiO_2}(s+d)}$$
$$a_{17}=-e^{+j\gamma_{SiO_2}(s+d)},a_{18}=e^{-j\gamma_{SiO_2}(s)},a_{19}=e^{+j\gamma_{SiO_2}(s)},a_{20}=-e^{-j\gamma_{Si}(s)},a_{21}=-e^{+j\gamma_{Si}(s)}$$

In (29), $f_{NL,z_1}, f_{NL,z_2}, g_{NL,z_1}, g_{NL,z_2}$ represent the following nonlinear functions:

$$f_{NL,z_1}=\text{sech}\left(\gamma_{NL}\left(s+d+t+z_1\right)\right)\times\tanh\left(\gamma_{NL}\left(s+d+t+z_1\right)\right) \tag{31}$$
$$f_{NL,z_2}=\text{sech}\left(\gamma_{NL}\left(z_2-s-d-t\right)\right)\times\tanh\left(\gamma_{NL}\left(z_2-s-d-t\right)\right) \tag{32}$$
$$g_{NL,z_1}=\text{sech}\left(\gamma_{NL}\left(s+d+t+z_1\right)\right) \tag{33}$$
$$g_{NL,z_2}=\text{sech}\left(\gamma_{NL}\left(z_2-s-d-t\right)\right) \tag{34}$$

Now, our analytical model for the proposed structure is completed. To solve the nonlinear system of equations given in (29), familiar techniques in the literature, such as Newton's method [100], Broyden's method [101], and the Finite-difference method [102], can be used. These methods have been defined in MATLAB software. However, the explanation of the solving procedure of the above equations is outside the scope of this article. In what follows, we will discuss and investigate the analytical results of the proposed model.

## 3. Results and Discussions

In this section, we will study the numerical results of the presented model to investigate the behavior of HSP[3]. To simulate the structure, the parameters of nonlinear layers are supposed to be: $\alpha = 6.4\times10^{-12} m^2 V^{-2}, \varepsilon_L = 2.405, \alpha' \left|H_{y,0}(0)\right|^2 = 0.5$ [97]. Without the lack of generality, we assume that both graphene sheets have similar parameters: $\mu_c = 0.35\ ev, \tau = 45\ fs, T = 300\ K, \Delta = 1nm.$ The permittivity of Si and $SiO_2$ layers are supposed to be $\varepsilon_{Si} = 11.9, \varepsilon_{SiO_2} = 2.09$, respectively. The parameters of hBN layers are given before. The configuration parameters of the structure are $t = 10nm,, d = 20nm, s = 30\ nm.$

Before embarking on a discussion about the numerical results, let us briefly review the permittivity function of hBN. As mentioned before, hBN is a popular anisotropic crystal with two phonon modes called "in-plane and out-of-plane phonon modes" [56]. These modes form two bands: 1- The lower Reststrahlen band related to type-I hyperbolicity, and 2- The upper Reststrahlen band related to type-II hyperbolicity [56]. In fig. 2, the real part of hBN permittivity has been depicted where two Reststrahlen bands have been highlighted. Now, we will study the propagating features of HSP[3] in the upper and lower Reststrahlen bands separately. It should be noted that our model is also valid once the losses of graphene sheets and hBN layers are included.

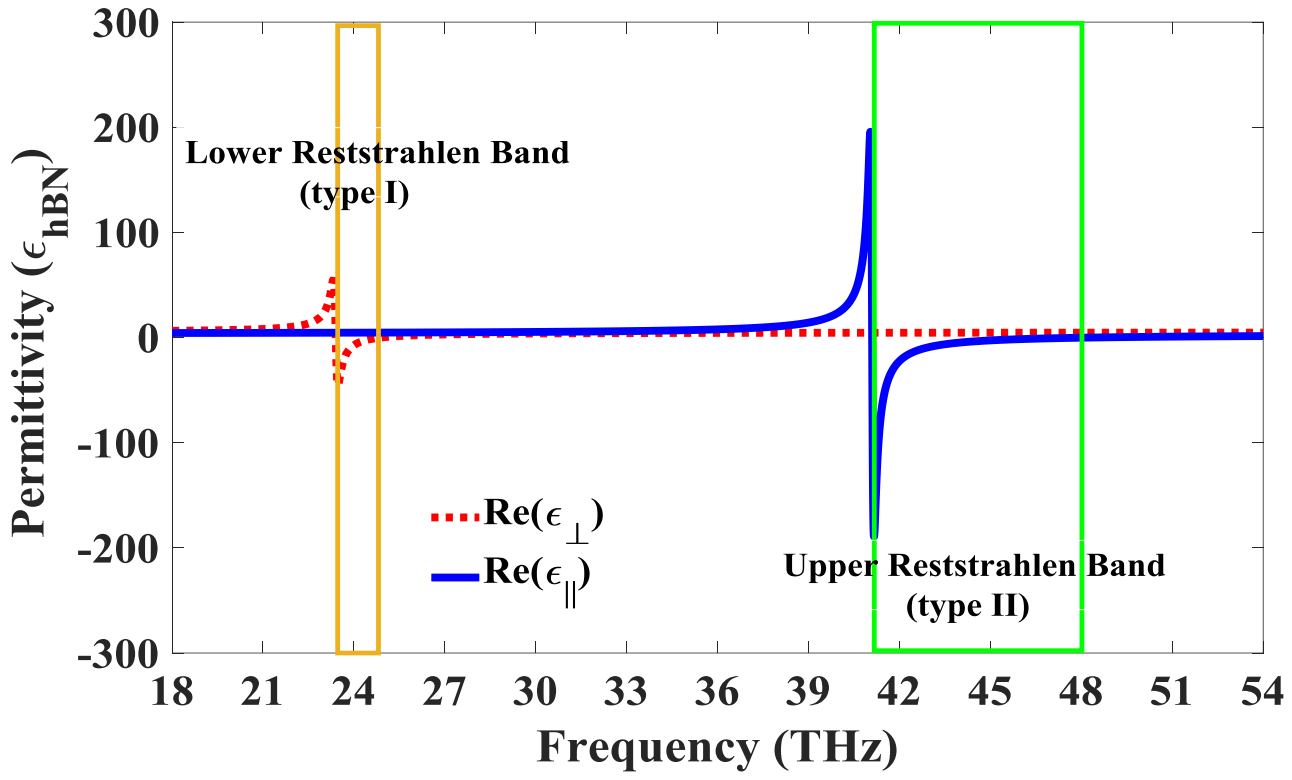

**Fig. 2.** The permittivity of hBN versus frequency. The lower and upper Reststrahlen bands are shown in this figure.

### 3.1. The Upper Reststrahlen Band

Fig. 3 illustrates the dependence of real and imaginary parts of propagation constant on the frequency in the upper Reststrahlen band. As seen in this figure, there are two hybrid modes outside the upper Reststrahlen band which we have called "Higher modes" and "Lower modes". The higher modes and lowers modes correspond to the high and low-frequency modes (with the same β), respectively. These modes exist due to the hBN layer (its origin is the permittivity tensor).

In Figures 4-6, the propagating properties of HSP[3] have been depicted as a function of the nonlinear coefficient for various values of the chemical potential. In these figures, the effective index and the propagation length are defined by $N_{eff} = Re[\beta]/k_0$, $L_{Prop} = 1/Im[\beta]$, respectively, where $k_0$ is the free-space wavenumber, and β is the propagation constant. Furthermore, the Figure of Merit (FOM) is defined as $FOM = Re[\beta]/Im[\beta]$ [103]. Moreover, all propagating properties have been shown at three specific frequencies: $f = 36\ THz$ (chosen in $\omega < \omega_{TO,\|}$), $f = 45\ THz$ (chosen in the upper Reststrahlen band, i.e. $\omega_{TO,\|} < \omega < \omega_{LO,\|}$), and $f = 54\ THz$ (chosen in $\omega > \omega_{LO,\|}$).

As observed in Fig. 4(a), the effective indices of lower modes increase with the increment of the nonlinear factor while they decrease slowly for the higher modes. By increasing the level of chemical potential, the effective index reduces, both for higher and lower modes. Moreover, it can be found that the difference between the two modes decreases as the chemical potential increases. It is clear from fig. 4(b) that the higher modes have a long propagation length, compared to those in lower modes. For both higher and lower modes, the increment of chemical doping will enhance the propagation length. The slope of propagation length variations is slow for the lower modes. One can see from fig. 4(c) that the FOM of the two modes is very close to each other, however, the higher modes have better FOM than the lower modes. At this frequency (36 THz), the FOM is not high due to the low propagation length.

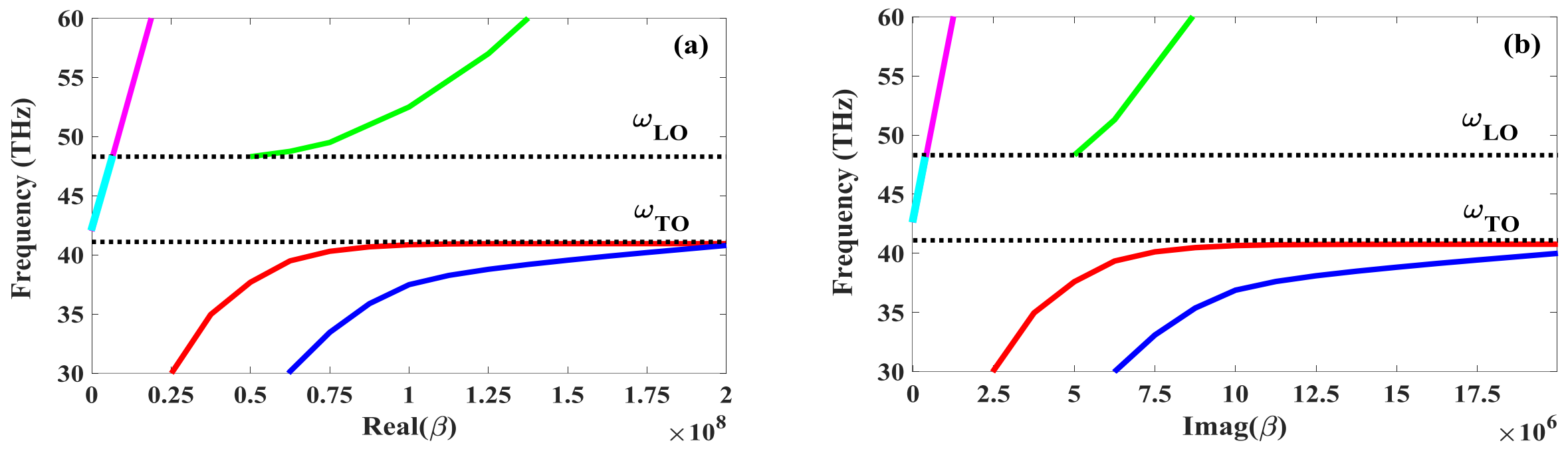

**Fig. 3.** The dispersion diagram of the proposed structure in the upper Reststrahlen band: **(a)** Real part of propagation constant versus frequency, **(b)** Imaginary part of propagation constant versus frequency. The nonlinear coefficient is $\alpha'|H_{y,0}|^2 = 0.5$, the chemical potential of graphene is 0.35 eV and the relaxation time is supposed to be 45 fs.

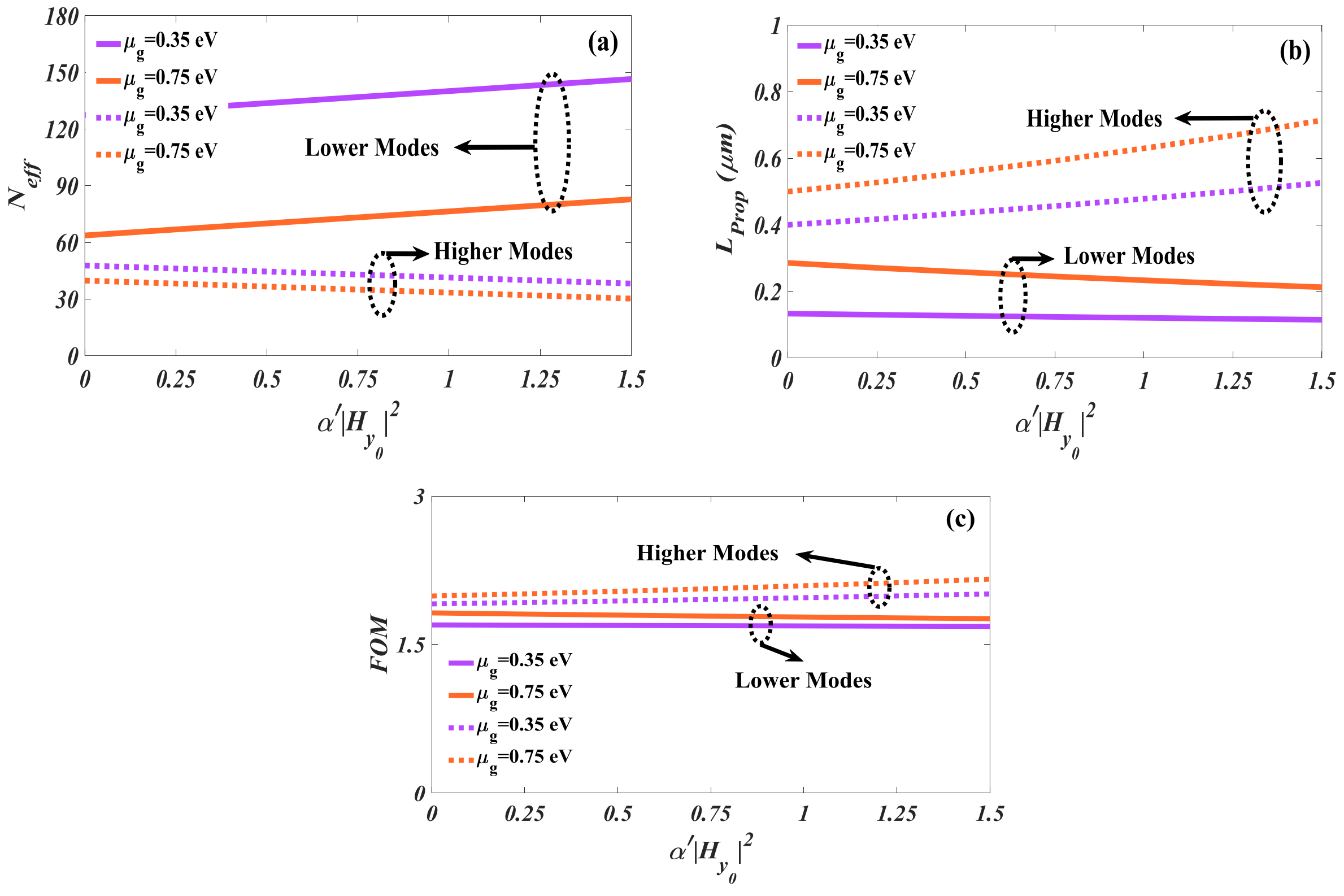

**Fig. 4.** The propagating features of the structure versus the nonlinear coefficient for various values of the chemical potential at the frequency of 36 THz: **(a)** the effective index, **(b)** the propagation length, **(c)** FOM. The relaxation time is supposed to be 45 fs. The solid and dashed lines represent the lower and higher modes, respectively.

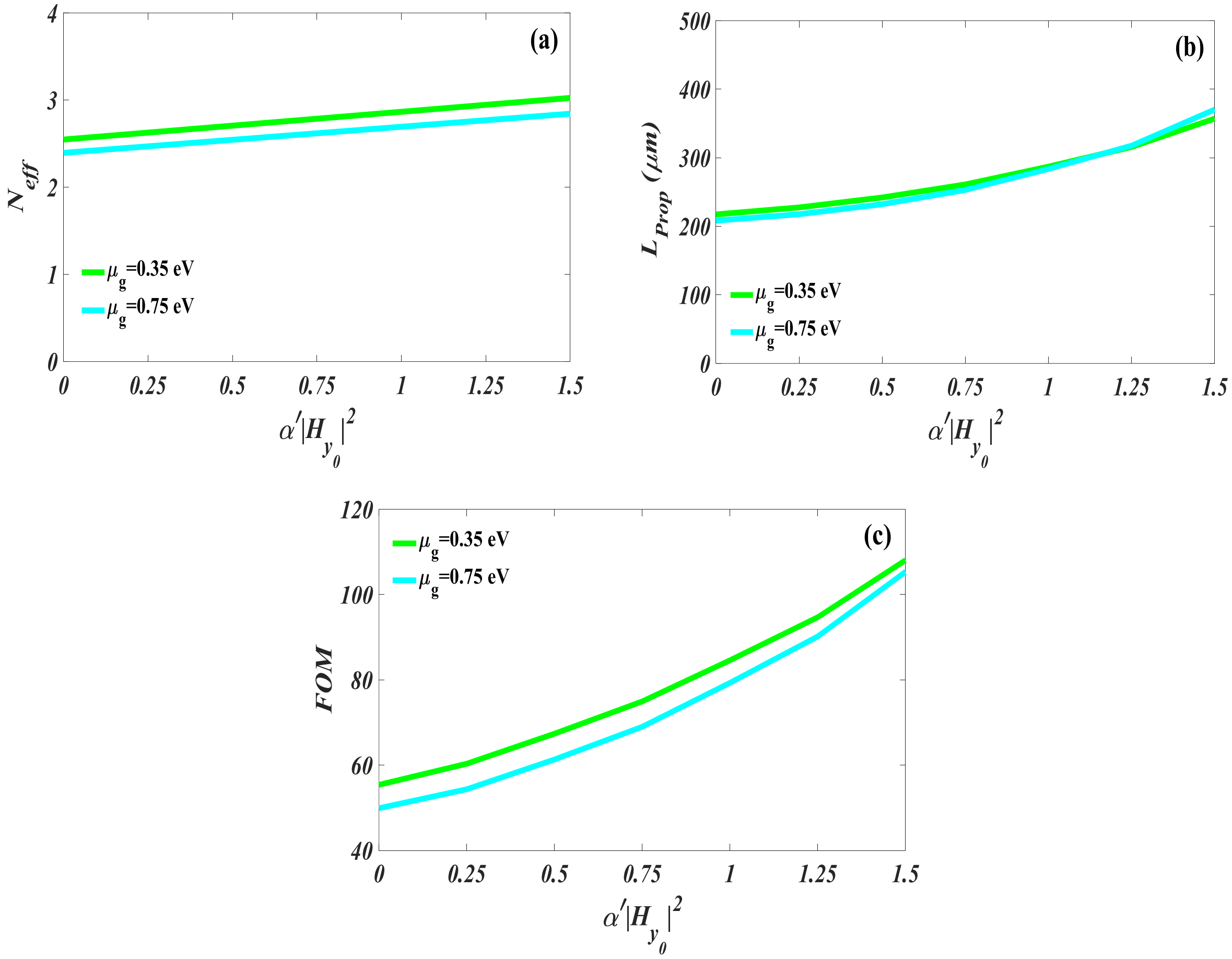

**Fig. 5.** The propagating features of the structure versus the nonlinear coefficient for various values of the chemical potential at the frequency of 45 THz: **(a)** the effective index, **(b)** the propagation length, **(c)** FOM. The relaxation time is 45 fs. The solid and dashed lines represent the lower and higher modes, respectively.

In fig. 5, the plasmonic features have been shown at 45 THz, a specific frequency chosen inside the upper Reststrahlen band. As explained before, there is only one propagating mode in this band, as confirmed in fig. 5. The increment of chemical potential reduces the effective index slightly, while it has a negligible effect on the propagation constant. Therefore, tuning the features of $HSP^3$ in this band only depends on the nonlinear factor. Compared to fig. 4(b), the propagation length is high in this frequency range, which causes higher values of FOM than those outside the upper Reststrahlen band. For instance, the propagation length and FOM of $HSP^3$ at this frequency (45 THz) reach 300μm and FOM=98, respectively, for the chemical potential of 0.35 eV and the nonlinear factor of $\alpha'\left|H_{y,0}\right|^2 = 1.25$. Our results for FOM and the propagation length at the frequency of 45 THz is a significant enhancement, compared to the previous works [104-107]. For instance, in [104], a quality factor of 33 and a propagation length of 3.2 μm are reported for propagating plasmons on a fabricated graphene-based structure. The authors in [105] have reported the FOM of 25 in a combined graphene-hBN structure. In [106], a FOM of 45 is achieved for the first mode of graphene-coated hBN nanowires at the frequency of 44.8 THz.

In the specific frequency chosen above the $\omega_{LO,\parallel}$, the effective indices of higher modes have a slight value, while lower modes have a better effective index, as seen in fig. 6. The propagation length has an opposite trend which means that higher modes have higher propagation length. Similar to fig. 4, the increment of the chemical doping will reduce the difference between the two hybrid modes. The values of FOM are low for these modes at the frequency of 54 THz because the propagation loss is so high (or the propagation length is low).

Let us consider what happens after placing graphene layers on our proposed structure. For instance, suppose the upper Reststrahlen band (similar explanations can be expressed for the lower Reststrahlen band). Before depositing graphene sheets, surface phonons only exist $\omega_{TO,\|} < \omega < \omega_{LO,\|}$ because of $\mathrm{Re}[\varepsilon_{\|}] < 0$. Now, by placing graphene sheets, the excited plasmons on graphene layers can be coupled to surface phonons on hBN, which now exist both inside and outside the Reststrahlen band. Due to the existence of surface phonons in the Reststrahlen band before depositing graphene, their coupling with surface plasmons will be stronger than those outside the Reststrahlen band. Thus, FOM is higher in the Reststrahlen band compared to outside of it, because the coupling of plasmons and phonons is strong.

As a final point for this sub-section, we investigate the effect of relaxation time on the FOM of $HSP^3$ at the frequencies of 36 THz, 45 THz, and 54 THz. It can be found from fig. 7 that the FOM of all propagating modes in these frequencies increases as the relaxation time of graphene sheets increases. At the frequency of 36 THz, the increment of relaxation time has a noticeable effect on the FOM of higher modes, compared to lower modes. Furthermore, the FOM of higher modes is more sensitive to the variations of nonlinearity for higher values of relaxation time ($\tau = 90\ fs$) at 36 THz. In the upper Reststrahlen band, fig. 7 (b) shows that the FOM increases as the relaxation time increases. For instance, FOM= 108 is achievable for the relaxation time of $\tau = 90\ fs,$ and the nonlinear factor of $\alpha' |H_{y,0}|^2 = 1.5$. As seen in fig. 7(c), the behavior of FOM variations is similar to Fig. 7(a).

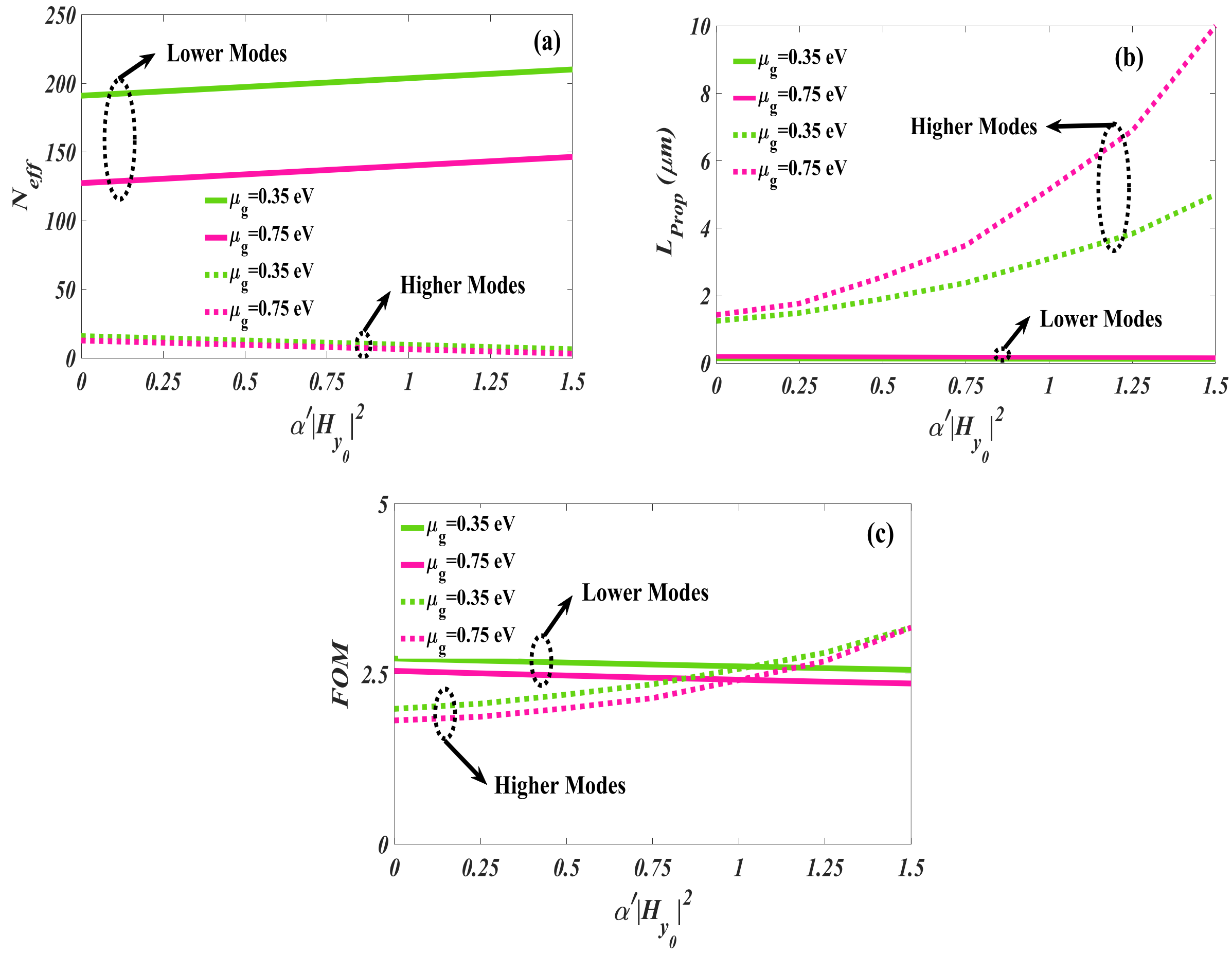

**Fig. 6.** The propagating features of the structure versus the nonlinear coefficient for various values of the chemical potential at the frequency of 54 THz: **(a)** the effective index, **(b)** the propagation length, **(c)** FOM. The relaxation time is 45 fs. The solid and dashed lines represent the lower and higher modes, respectively.

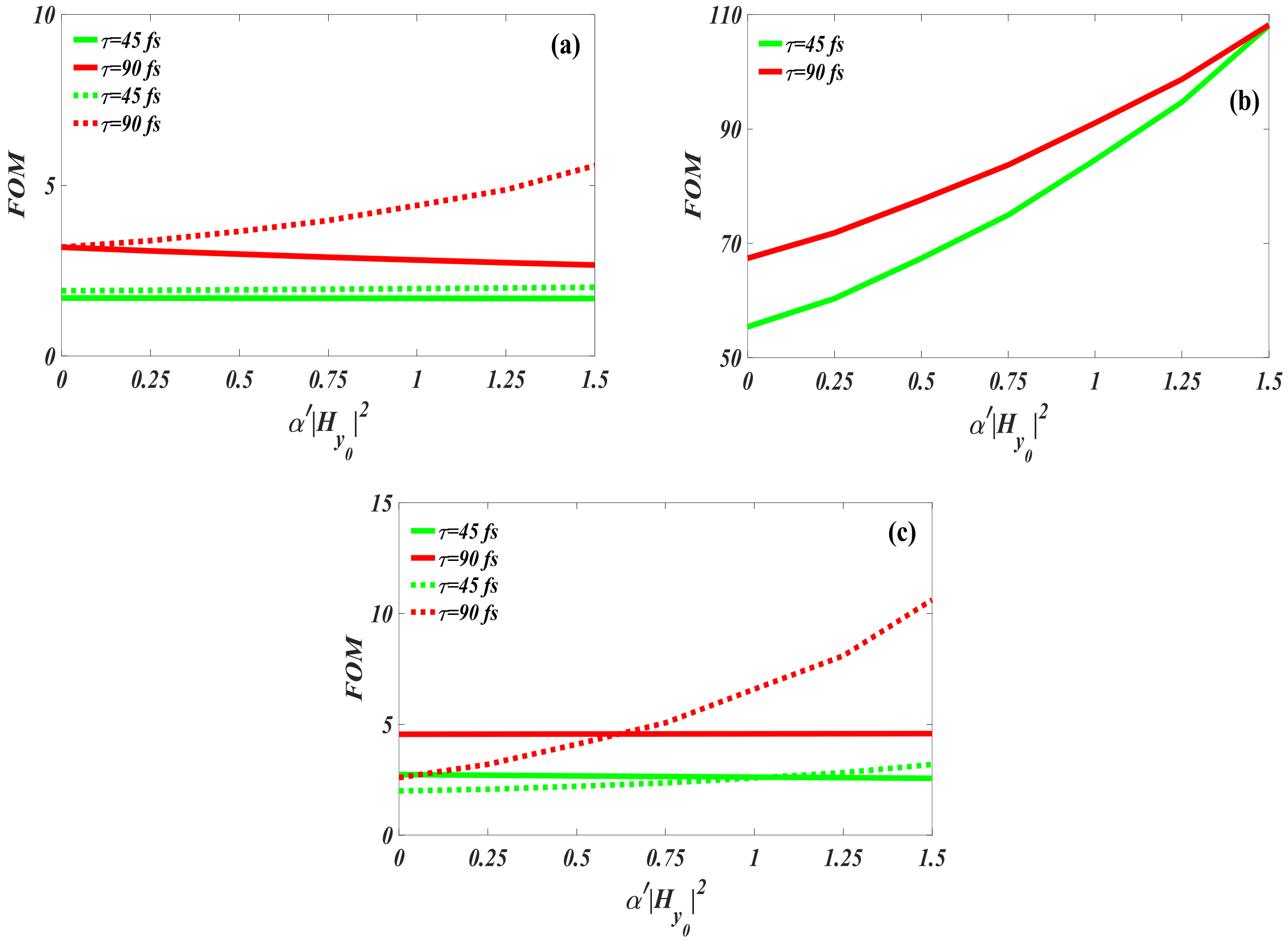

**Fig. 7.** The dependence of FOM on the nonlinear coefficient for various values of the relaxation time at the frequency of: **(a)** 36 THz, **(b)** 45 THz, **(c)** 54 THz. The chemical potential is supposed to be 0.35 eV. The solid and dashed lines represent the lower and higher modes, respectively.

### 3.2. The Lower Reststrahlen Band

In the previous section, we considered the numerical results in the upper Reststrahlen band. Here, we will investigate the behavior of $HSP^3$ in the lower Reststrahlen band. It should be noted that all parameters (except the studied frequency range) such as configuration parameters, nonlinear coefficients, etc. are remained fixed unless otherwise stated.

Fig. 8 demonstrates the dispersion diagram of the structure in the lower Reststrahlen band. Similar to fig. 3, it is obvious that two hybrid modes exist outside the lower Reststrahlen band, while there is only one mode inside this band.

In figures 9-10, the dependence of propagating features on the nonlinear factor has been represented for various values of the chemical potential at the two specific frequencies: 21 THz (chosen below the $\omega_{TO,\perp}$) and 24 THz (chosen in the lower Reststrahlen band). We should note that the plasmonic properties have not been depicted for a specific frequency in the range of $\omega > \omega_{LO,\perp}$ because the features of $HSP^3$ are the same for $\omega > \omega_{LO,\perp}$ and $\omega < \omega_{LO,\parallel}$ (the study has been done for the frequency of 36 THz in fig. 4 in the previous sub-section); so that we do not repeat those results.

As observed in fig. 9(a), the effective index of lower modes increases as the nonlinear coefficient increases while they have the opposite trend for higher modes. The difference between the two hybrid modes decreases with the increment of the chemical potential. One can see from fig. 9(b) that higher modes have longer propagation lengths in comparison with lower modes. Thus, the FOM of $HSP^3$ for lower modes is higher than those for higher modes. Another interesting point is the independence of the FOM of lower modes on the chemical potential variations.

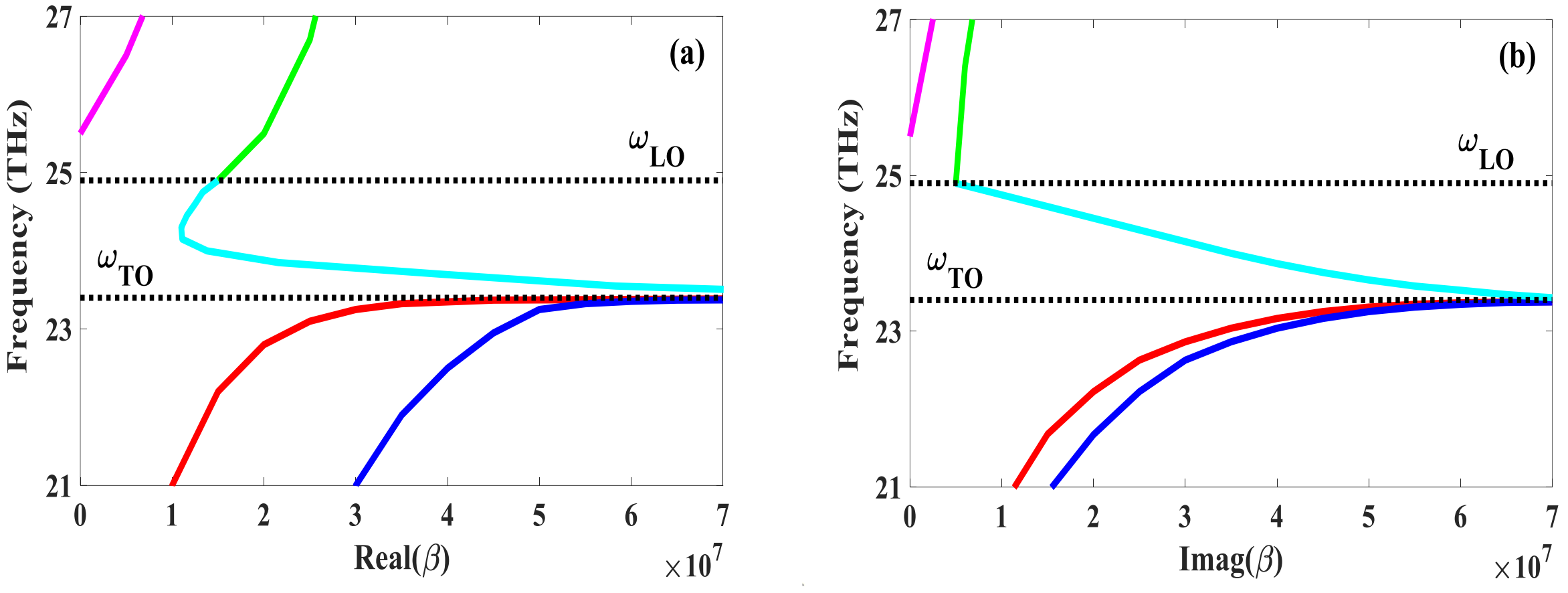

**Fig. 8.** The dispersion diagram of the proposed structure in the lower Reststrahlen band: **(a)** Real part of propagation constant versus frequency**, (b)** Imaginary part of propagation constant versus frequency. The nonlinear coefficient is $\alpha' |H_{y,0}|^2 = 0.5$, the chemical potential of graphene is 0.35 eV and the relaxation time is supposed to be 45 fs.

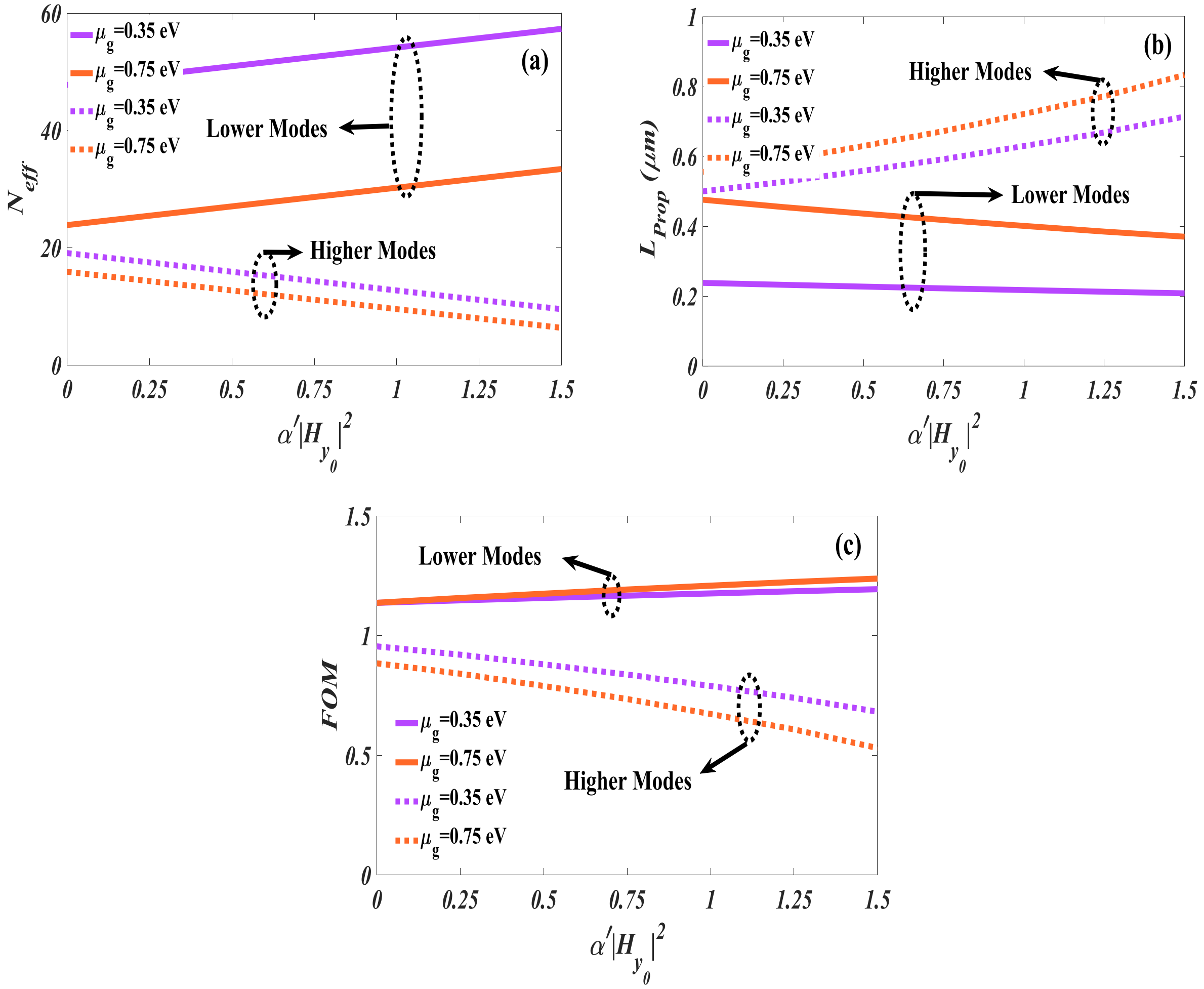

**Fig. 9.** The propagating features of the structure versus the nonlinear coefficient for various values of the chemical potential at the frequency of 21 THz: (**a**) the effective index, **(b)** the propagation length, **(c)** FOM. The relaxation time is 45 fs. The solid and dashed lines represent the lower and higher modes, respectively.

It can be seen from fig. 10 that the effective index and propagation length have opposite trends by the increment of the nonlinear factor, which means that the effective index increases as the nonlinear coefficient increases while the propagation length decreases (the slope of variations for the propagation length of $\mu_g = 0.35\ ev$ is slight). For a fixed nonlinear coefficient, a higher value of chemical potential results in a lower effective index and longer propagation length. Compared to the upper Reststrahlen band, FOM has an opposite trend for higher chemical potentials (compare fig. 10(c) with fig. 5(c)). Moreover, FOM has higher values in the upper Reststrahlen band than that in the lower Reststrahlen band.

The effect of relaxation time on the performance of $HSP^3$ is depicted in fig. 11 at the frequencies of 21 THz and 24 THz. As seen in this figure, higher relaxation time gives high FOM, both for two modes at 21 THz. However, at the frequency of 24 THz, FOM increases slightly as the relaxation time increases.

As a final point, we investigate the influence of hBN thickness on FOM in the upper and lower Reststrahlen band. One can observe from fig. 12 that as the thickness of hBN increases from 10nm to 20nm, FOM at the frequency of 45 THz increases from 80 to 88. At the frequency of 24 THz (in the lower Reststrahlen band), as the thickness of hBN increases, FOM increases more and more slowly.

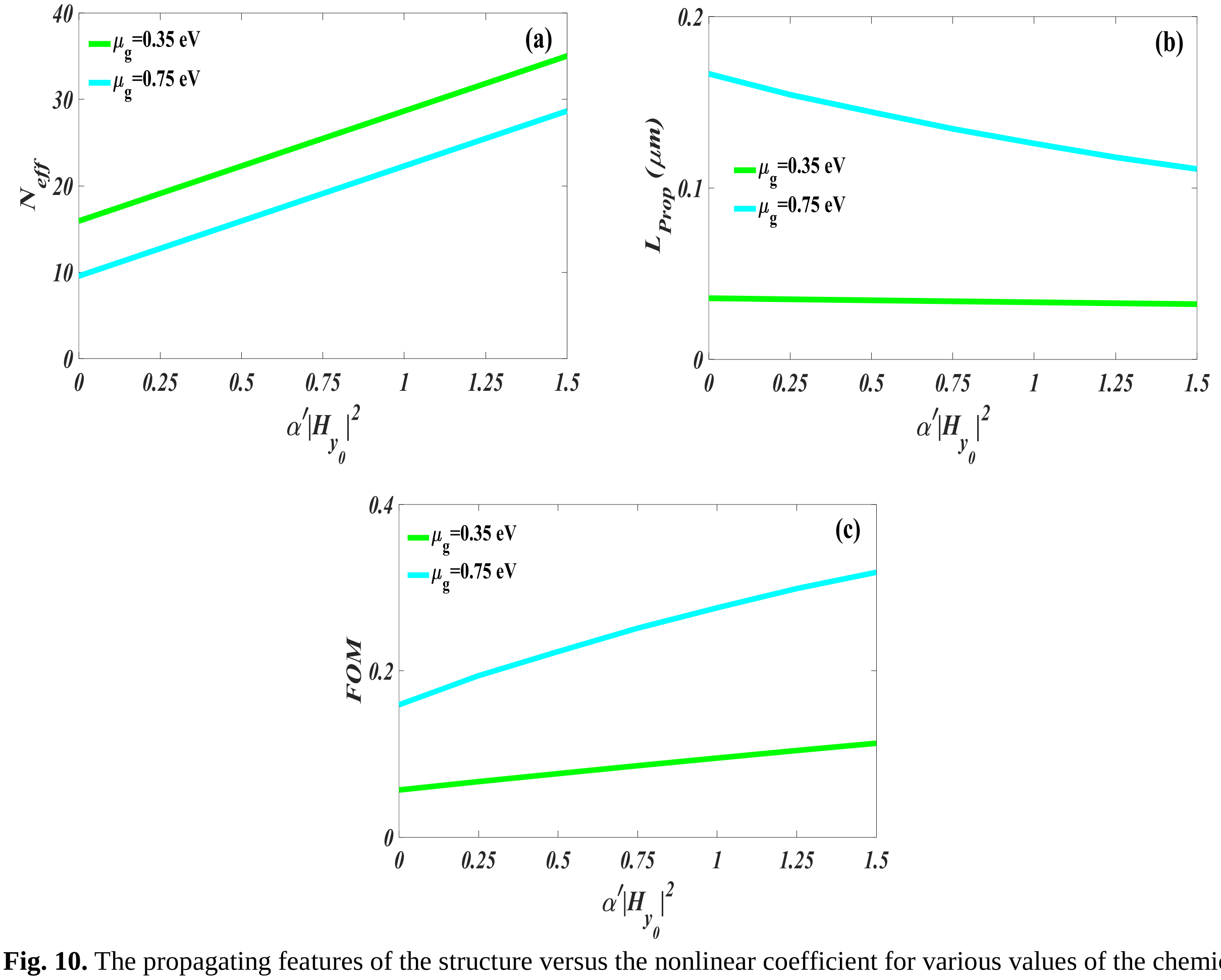

**Fig. 10.** The propagating features of the structure versus the nonlinear coefficient for various values of the chemical potential at the frequency of 24 THz: (**a**) the effective index, **(b)** the propagation length, **(c)** FOM. The relaxation time is 45 fs. The solid and dashed lines represent the lower and higher modes, respectively.

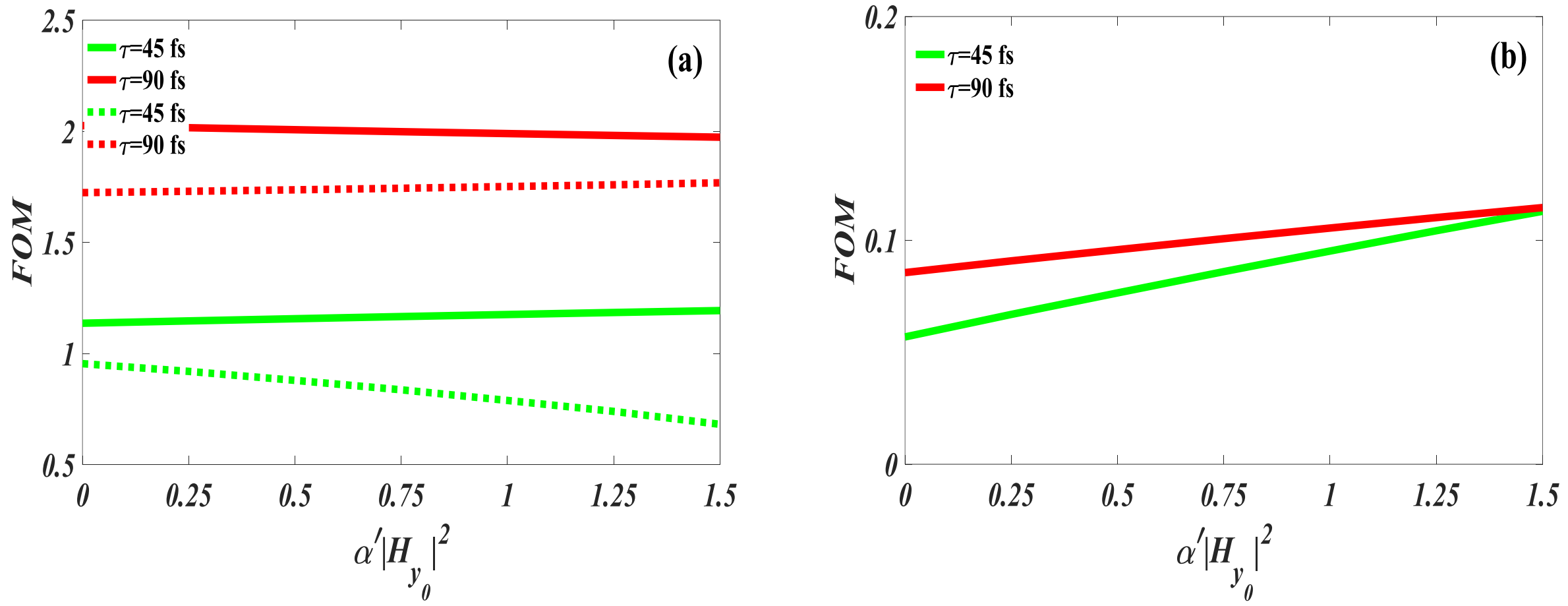

**Fig. 11.** The dependence of FOM on the nonlinear coefficient for various values of the relaxation time at the frequency of: **(a)** 21 THz, **(b)** 24 THz. The chemical potential is supposed to be 0.35 eV. The solid and dashed lines represent the lower and higher modes, respectively.

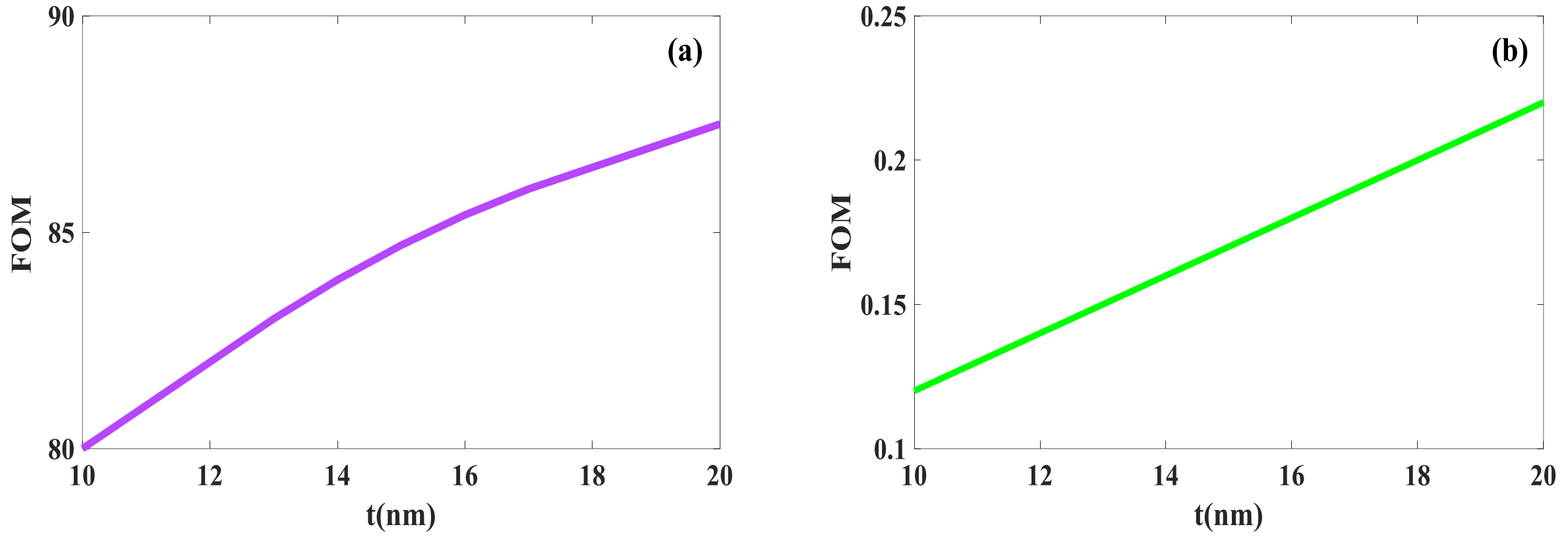

**Fig. 12.** The dependence of FOM on the thickness of hBN at the frequency of: **(a)** 45 THz, **(b)** 24 THz. The chemical potential is supposed to be 0.35 eV. The nonlinear coefficient is $\alpha'\left|H_{y,0}\right|^2 = 0.75$.

## 4. Conclusion

In this paper, the analytical expressions were derived for the study of $HSP^3$ in a nonlinear graphene-based structure with hBN layers. The numerical results showed that the propagation features of $HSP^3$ can be varied by changing the chemical potential, the relaxation time, and the nonlinear coefficient. Due to the existence of hBN layers in the proposed structure, the investigation was done in two frequency ranges: the upper and lower Reststrahlen bands. The obtained results represented that two kinds of hybrid modes propagate outside the upper and lower Reststrahlen bands, which we called “higher and lower modes”. However, only one propagating mode was seen inside the upper or lower Reststrahlen band. A high value of FOM=98 was reported at the frequency of 45 THz (inside the upper Reststrahlen band) for the chemical potential of 0.35 eV. The authors believe that the analytical study done on $HSP^3$ in this paper can help the researchers to design innovative structures such as sensors in the mid-infrared region.

## Declarations

**Ethics Approval:** Not Applicable.

**Consent to Participate:** Not Applicable.

**Consent for Publication:** Not Applicable.

**Funding:** The authors received no specific funding for this work.

**Conflicts of Interest/ Competing Interests:** The authors declare no competing interests.

**Availability of Data and Materials:** Not Applicable.

**Code availability:** Not Applicable.

**Authors' Contributions:** M. B. Heydari proposed the main idea of this work and performed the analytical modeling. M. Karimipour conducted the numerical simulations and wrote the manuscript. M. Mohammadi Shirkolaei analyzed the results and reviewed the paper.

## References

[1] S. A. Maier, *Plasmonics: fundamentals and applications*: Springer Science & Business Media, 2007.

[2] E. Ozbay, "Plasmonics: merging photonics and electronics at nanoscale dimensions," *science,* vol. 311, pp. 189-193, 2006.

[3] S. Dai, Z. Fei, Q. Ma, A. Rodin, M. Wagner, A. McLeod, *et al.*, "Tunable phonon polaritons in atomically thin van der Waals crystals of boron nitride," *Science,* vol. 343, pp. 1125-1129, 2014.

[4] J. D. Caldwell, A. V. Kretinin, Y. Chen, V. Giannini, M. M. Fogler, Y. Francescato, *et al.*, "Sub-diffractional volume-confined polaritons in the natural hyperbolic material hexagonal boron nitride," *Nature communications,* vol. 5, pp. 1-9, 2014.

[5] M. Kafesaki, A. A. Basharin, E. N. Economou, and C. M. Soukoulis, "THz metamaterials made of phonon-polariton materials," *Photonics and Nanostructures - Fundamentals and Applications,* vol. 12, pp. 376-386, 2014/08/01/ 2014.

[6] M. Zhai, H. Peng, X. Wang, X. Wang, Z. Chen, and W. Yin, "The Conformal HIE-FDTD Method for Simulating Tunable Graphene-Based Couplers for THz Applications," *IEEE Transactions on Terahertz Science and Technology,* vol. 5, pp. 368-376, 2015.

[7] M. B. Heydari and M. H. V. Samiei, "Graphene-Based Couplers: A Brief Review," *arXiv preprint arXiv:2010.09462,* 2020.

[8] F. Xu, H. Zhang, and Y. Sun, "Compact graphene directional couplers based on dielectric ridges for planar integration," *Optik-International Journal for Light and Electron Optics,* vol. 131, pp. 588-591, 2017.

[9] D. Correas-Serrano, J. S. Gomez-Diaz, J. Perruisseau-Carrier, and A. Álvarez-Melcón, "Graphene-Based Plasmonic Tunable Low-Pass Filters in the Terahertz Band," *IEEE Transactions on Nanotechnology,* vol. 13, pp. 1145-1153, 2014.

[10] M. B. Heydari and M. H. V. Samiei, "A Short Review on Graphene-Based Filters: Perspectives and Challenges," *arXiv preprint arXiv:2010.07176,* 2020.

[11] H. Zhuang, F. Kong, K. Li, and S. Sheng, "Plasmonic bandpass filter based on graphene nanoribbon," *Applied optics,* vol. 54, pp. 2558-2564, 2015.

[12] T. Christopoulos, O. Tsilipakos, N. Grivas, and E. E. Kriezis, "Coupled-mode-theory framework for nonlinear resonators comprising graphene," *Physical Review E,* vol. 94, p. 062219, 2016.

[13] M. B. Heydari and M. H. V. Samiei, "A Short Review of Plasmonic Graphene-Based Resonators: Recent Advances and Prospects," *arXiv preprint arXiv:2011.14767,* 2020.

[14] X. Zhou, T. Zhang, X. Yin, L. Chen, and X. Li, "Dynamically Tunable Electromagnetically Induced Transparency in Graphene-Based Coupled Micro-ring Resonators," *IEEE Photonics Journal,* vol. 9, pp. 1-9, 2017.

[15] V. Dmitriev, S. L. M. da Silva, and W. Castro, "Ultrawideband graphene three-port circulator for THz region," *Optics express,* vol. 27, pp. 15982-15995, 2019.

[16] M. B. Heydari and M. H. V. Samiei, "Three-port Terahertz Circulator with Multi-layer Triangular Graphene-Based Post," *Optik*, p. 166457, 2021.
[17] V. Dmitriev and W. Castro, "Dynamically controllable terahertz graphene Y-circulator," *IEEE Transactions on Magnetics*, vol. 55, pp. 1-12, 2018.
[18] M. B. Heydari and M. H. V. Samiei, "A Novel Graphene-Based Circulator with Multi-layer Triangular Post for THz Region," *arXiv preprint arXiv:2102.02683*, 2021.
[19] M. B. Heydari, "Highly Confined Mid-infrared Plasmonics in Graphene-plasma Cylindrical Structures," 2022.
[20] J. Christensen, A. Manjavacas, S. Thongrattanasiri, F. H. Koppens, and F. J. García de Abajo, "Graphene plasmon waveguiding and hybridization in individual and paired nanoribbons," *ACS nano*, vol. 6, pp. 431-440, 2012.
[21] G. W. Hanson, "Quasi-transverse electromagnetic modes supported by a graphene parallel-plate waveguide," *Journal of Applied Physics*, vol. 104, p. 084314, 2008.
[22] M. B. Heydari and M. H. V. Samiei, "A Novel Analytical Study of Anisotropic Multi-Layer Elliptical Structures Containing Graphene Layers," *IEEE Transactions on Magnetics*, vol. 56, pp. 1-10, 2020.
[23] M. B. Heydari and M. H. V. Samiei, "Analytical Study of Chiral Multi-Layer Structures Containing Graphene Sheets for THz Applications," *IEEE Transactions on Nanotechnology*, vol. 19, pp. 653-660, 2020.
[24] K. Jamalpoor, A. Zarifkar, and M. Miri, "Application of graphene second-order nonlinearity in THz plasmons excitation," *Photonics and Nanostructures - Fundamentals and Applications*, vol. 26, pp. 80-84, 2017/09/01/ 2017.
[25] S. Khoubafarin Doust, V. Siahpoush, and A. Asgari, "The Tunability of Surface Plasmon Polaritons in Graphene Waveguide Structures," *Plasmonics*, vol. 12, pp. 1633-1639, 2017/10/01 2017.
[26] C. E. Talley, J. B. Jackson, C. Oubre, N. K. Grady, C. W. Hollars, S. M. Lane*, et al.*, "Surface-enhanced Raman scattering from individual Au nanoparticles and nanoparticle dimer substrates," *Nano letters*, vol. 5, pp. 1569-1574, 2005.
[27] M. Saifur Rahman, K. A. Rikta, L. F. Abdulrazak, and M. S. Anower, "Enhanced performance of SnSe-Graphene hybrid photonic surface plasmon refractive sensor for biosensing applications," *Photonics and Nanostructures - Fundamentals and Applications*, vol. 39, p. 100779, 2020/05/01/ 2020.
[28] M. Uwais, A. Bijalwan, and V. Rastogi, "Surface Phonon Resonance Assisted Refractive Index Sensing in Mid-Infrared Wavelength Range," *Plasmonics*, 2022/03/28 2022.
[29] D. Rodrigo, O. Limaj, D. Janner, D. Etezadi, F. J. García de Abajo, V. Pruneri*, et al.*, "Mid-infrared plasmonic biosensing with graphene," *Science*, vol. 349, pp. 165-168, 2015.
[30] H. Hu, X. Yang, F. Zhai, D. Hu, R. Liu, K. Liu*, et al.*, "Far-field nanoscale infrared spectroscopy of vibrational fingerprints of molecules with graphene plasmons," *Nature communications*, vol. 7, pp. 1-8, 2016.
[31] H. Hu, X. Yang, X. Guo, K. Khaliji, S. R. Biswas, F. J. García de Abajo*, et al.*, "Gas identification with graphene plasmons," *Nature communications*, vol. 10, pp. 1-7, 2019.
[32] H. Hajian, H. Caglayan, and E. Ozbay, "Long-range Tamm surface plasmons supported by graphene-dielectric metamaterials," *Journal of Applied Physics*, vol. 121, p. 033101, 2017.
[33] N. Fang, H. Lee, C. Sun, and X. Zhang, "Sub-diffraction-limited optical imaging with a silver superlens," *science*, vol. 308, pp. 534-537, 2005.
[34] R. E. da Silva, R. Macêdo, T. Dumelow, J. Da Costa, S. Honorato, and A. Ayala, "Far-infrared slab lensing and subwavelength imaging in crystal quartz," *Physical Review B*, vol. 86, p. 155152, 2012.
[35] M. B. Heydari and M. H. V. Samiei, "Plasmonic graphene waveguides: A literature review," *arXiv preprint arXiv:1809.09937*, 2018.
[36] W. Fuscaldo, P. Burghignoli, P. Baccarelli, and A. Galli, "Complex mode spectra of graphene-based planar structures for THz applications," *Journal of Infrared, Millimeter, and Terahertz Waves*, vol. 36, pp. 720-733, 2015.

[37] M. B. Heydari and M. H. Vadjed Samiei, "An Analytical Study of Magneto-Plasmons in Anisotropic Multi-layer Structures Containing Magnetically Biased Graphene Sheets," *Plasmonics*, vol. 15, pp. 1183-1198, 2020/08/01 2020.
[38] D. Correas-Serrano, J. S. Gomez-Diaz, J. Perruisseau-Carrier, and A. Álvarez-Melcón, "Spatially dispersive graphene single and parallel plate waveguides: Analysis and circuit model," *IEEE Transactions on Microwave Theory and Techniques,* vol. 61, pp. 4333-4344, 2013.
[39] M. B. Heydari and M. H. Vadjed Samiei, "New analytical investigation of anisotropic graphene nano-waveguides with bi-gyrotropic cover and substrate backed by a PEMC layer," *Optical and Quantum Electronics*, vol. 52, p. 108, 2020/02/07 2020.
[40] M. B. Heydari and M. H. V. Samiei, "Analytical Investigation of Magneto-Plasmons in Anisotropic Multi-layer Planar Waveguides Incorporating Magnetically Biased Graphene Sheets," *arXiv preprint arXiv:2103.11452,* 2021.
[41] I. D. Koufogiannis, M. Mattes, and J. R. Mosig, "On the development and evaluation of spatial-domain Green's functions for multilayered structures with conductive sheets," *IEEE Transactions on Microwave Theory and Techniques,* vol. 63, pp. 20-29, 2015.
[42] M. B. Heydari and M. H. V. Samiei, "Magneto-Plasmons in Grounded Graphene-Based Structures with Anisotropic Cover and Substrate," *arXiv preprint arXiv:2103.08557*, 2021.
[43] M. B. Heydari, "Hybrid Graphene-Gyroelectric Structures: A Novel Platform for THz Applications," *arXiv preprint arXiv:2201.06538,* 2022.
[44] M. B. Heydari, "Tunable Plasmonic Modes in Graphene-loaded Plasma Media," 2022.
[45] Y. T. Aladadi and M. A. Alkanhal, "Electromagnetic Characterization of Graphene-Plasma Formations," *IEEE Transactions on Plasma Science,* vol. 48, pp. 852-857, 2020.
[46] Y. Gao, G. Ren, B. Zhu, J. Wang, and S. Jian, "Single-mode graphene-coated nanowire plasmonic waveguide," *Optics letters,* vol. 39, pp. 5909-5912, 2014.
[47] Y. Gao, G. Ren, B. Zhu, H. Liu, Y. Lian, and S. Jian, "Analytical model for plasmon modes in graphene-coated nanowire," *Optics express,* vol. 22, pp. 24322-24331, 2014.
[48] M. B. Heydari and M. H. V. Samiei, "Novel analytical model of anisotropic multi-layer cylindrical waveguides incorporating graphene layers," *Photonics and Nanostructures-Fundamentals and Applications,* vol. 42, p. 100834, 2020.
[49] D. A. Kuzmin, I. V. Bychkov, V. G. Shavrov, and L. N. Kotov, "Transverse-electric plasmonic modes of cylindrical graphene-based waveguide at near-infrared and visible frequencies," *Scientific reports,* vol. 6, p. 26915, 2016.
[50] M. B. Heydari and M. H. V. Samiei, "Anisotropic Multi-layer Cylindrical Structures Containing Graphene Layers: An Analytical Approach," *arXiv preprint arXiv:2103.05594,* 2021.
[51] M. B. Heydari, "Tunable SPPs in graphene-based cylindrical structures with gyroelectric layers," *Optik,* vol. 254, p. 168651, 2022.
[52] M. B. Heydari, "Novel Theoretical Study of Plasmonic Waves in Graphene-based Cylindrical Waveguides with Gyro-electric Layers," *arXiv preprint arXiv:2201.10421,* 2022.
[53] D. Teng, K. Wang, Z. Li, Y. Zhao, G. Zhao, H. Li*, et al.*, "Graphene-Coated Elliptical Nanowires for Low Loss Subwavelength Terahertz Transmission," *Applied Sciences,* vol. 9, p. 2351, 2019.
[54] X. Cheng, W. Xue, Z. Wei, H. Dong, and C. Li, "Mode analysis of a confocal elliptical dielectric nanowire coated with double-layer graphene," *Optics Communications,* vol. 452, pp. 467-475, 2019.
[55] M. B. Heydari and M. H. V. Samiei, "Anisotropic Multi-layer Elliptical Waveguides Incorporating Graphene Layers: A Novel Analytical Model," *arXiv preprint arXiv:2103.01925,* 2021.
[56] A. Kumar, T. Low, K. H. Fung, P. Avouris, and N. X. Fang, "Tunable light–matter interaction and the role of hyperbolicity in graphene–hBN system," *Nano letters,* vol. 15, pp. 3172-3180, 2015.
[57] E. L. Wolf, *Applications of graphene: an overview*: Springer Science & Business Media, 2014.
[58] P. Tassin, T. Koschny, and C. M. Soukoulis, "Graphene for terahertz applications," *Science,* vol. 341, pp. 620-621, 2013.

[59] M. B. Heydari, "Tunable SPPs supported by hybrid graphene-gyroelectric waveguides: an analytical approach," *Optical and Quantum Electronics,* vol. 54, pp. 1-14, 2022.
[60] M. Z. Yaqoob, A. Ghaffar, M. Alkanhal, and S. U. Rehman, "Characteristics of light–plasmon coupling on chiral–graphene interface," *JOSA B,* vol. 36, 2019// 2019.
[61] M. Yaqoob, A. Ghaffar, M. Alkanhal, S. ur Rehman, and F. Razzaz, "Hybrid Surface Plasmon Polariton Wave Generation and Modulation by Chiral-Graphene-Metal (CGM) Structure," *Scientific reports,* vol. 8, pp. 1-9, 2018.
[62] M. B. Heydari and M. H. V. Samiei, "Chiral Multi-layer Waveguides Incorporating Graphene Sheets: An Analytical Approach," *arXiv preprint arXiv:2102.10135,* 2021.
[63] I. Toqeer, A. Ghaffar, M. Y. Naz, and B. Sultana, "Characteristics of dispersion modes supported by Graphene Chiral Graphene waveguide," *Optik,* vol. 186, 2019// 2019.
[64] M. B. Heydari and M. H. Vadjed Samiei, "Analytical study of hybrid surface plasmon polaritons in a grounded chiral slab waveguide covered with graphene sheet," *Optical and Quantum Electronics,* vol. 52, p. 406, 2020/09/08 2020.
[65] R. Zhao, J. Li, Q. Zhang, X. Liu, and Y. Zhang, "Behavior of SPPs in chiral–graphene–chiral structure," *Optics Letters,* vol. 46, pp. 1975-1978, 2021.
[66] M. B. Heydari and M. H. V. Samiei, "Grounded Graphene-Based Nano-Waveguide with Chiral Cover and Substrate: New Theoretical Investigation," *arXiv preprint arXiv:2102.12465,* 2021.
[67] C. Bhagyaraj, R. Ajith, and M. Vincent, "Propagation characteristics of surface plasmon polariton modes in graphene layer with nonlinear magnetic cladding," *Journal of Optics,* vol. 19, p. 035002, 2017.
[68] M. B. Heydari and M. H. V. Samiei, "Analytical study of TM-polarized surface plasmon polaritons in nonlinear multi-layer graphene-based waveguides," *Plasmonics,* vol. 16, pp. 841-848, 2021.
[69] X. Jiang, J. Gao, and X. Sun, "Control of dispersion properties in a nonlinear dielectric-graphene plasmonic waveguide," *Physica E: Low-dimensional Systems and Nanostructures,* vol. 106, pp. 176-179, 2019.
[70] M. B. Heydari and M. H. V. Samiei, "TM-polarized Surface Plasmon Polaritons in Nonlinear Multi-layer Graphene-Based Waveguides: An Analytical Study," *arXiv preprint arXiv:2101.02536,* 2021.
[71] Y. V. Bludov, D. A. Smirnova, Y. S. Kivshar, N. Peres, and M. I. Vasilevskiy, "Nonlinear TE-polarized surface polaritons on graphene," *Physical Review B,* vol. 89, p. 035406, 2014.
[72] M. B. Heydari, "Analytical Study of TE-Polarized SPPs in Nonlinear Multi-Layer Graphene-Based Structures," *Plasmonics,* vol. 16, pp. 2327-2334, 2021.
[73] Y. Wu, X. Dai, Y. Xiang, and D. Fan, "Nonlinear TE-polarized SPPs on a graphene cladded parallel plate waveguide," *Journal of Applied Physics,* vol. 121, p. 103103, 2017.
[74] M. B. Heydari, "TE-Polarized Surface Plasmon Polaritons (SPPs) in Nonlinear Multi-layer Graphene-Based Waveguides: An Analytical Model," *arXiv preprint arXiv:2107.01684,* 2021.
[75] S. Baher and Z. Lorestaniweiss, "Propagation of surface plasmon polaritons in monolayer graphene surrounded by nonlinear dielectric media," *Journal of Applied Physics,* vol. 124, p. 073103, 2018.
[76] M. B. Heydari, "TE-Polarized SPPs in Nonlinear Multi-layer Graphene-Based Structures," 2021.
[77] W. Walasik, A. Rodriguez, and G. Renversez, "Symmetric plasmonic slot waveguides with a nonlinear dielectric core: Bifurcations, size effects, and higher order modes," *Plasmonics,* vol. 10, pp. 33-38, 2015.
[78] H. Hajian, A. Soltani-Vala, M. Kalafi, and P. T. Leung, "Surface plasmons of a graphene parallel plate waveguide bounded by Kerr-type nonlinear media," *Journal of Applied Physics,* vol. 115, p. 083104, 2014.
[79] H. Hajian, I. D. Rukhlenko, P. T. Leung, H. Caglayan, and E. Ozbay, "Guided plasmon modes of a graphene-coated Kerr slab," *Plasmonics,* vol. 11, pp. 735-741, 2016.
[80] L. Wang, W. Cai, X. Zhang, and J. Xu, "Surface plasmons at the interface between graphene and Kerr-type nonlinear media," *Optics letters,* vol. 37, pp. 2730-2732, 2012.
[81] Y. Wu, L. Jiang, H. Xu, X. Dai, Y. Xiang, and D. Fan, "Hybrid nonlinear surface-phonon-plasmon-polaritons at the interface of nolinear medium and graphene-covered hexagonal boron nitride crystal," *Optics Express,* vol. 24, pp. 2109-2124, 2016.

[82] N. Ocelic and R. Hillenbrand, "Subwavelength-scale tailoring of surface phonon polaritons by focused ion-beam implantation," *Nature materials,* vol. 3, pp. 606-609, 2004.
[83] T. Taubner, D. Korobkin, Y. Urzhumov, G. Shvets, and R. Hillenbrand, "Near-field microscopy through a SiC superlens," *Science,* vol. 313, pp. 1595-1595, 2006.
[84] P. A. Belov, "Backward waves and negative refraction in uniaxial dielectrics with negative dielectric permittivity along the anisotropy axis," *Microwave and Optical Technology Letters,* vol. 37, pp. 259-263, 2003.
[85] R. Koch, T. Seyller, and J. Schaefer, "Strong phonon-plasmon coupled modes in the graphene/silicon carbide heterosystem," *Physical Review B,* vol. 82, p. 201413, 2010.
[86] T. Marinković, I. Radović, D. Borka, and Z. L. Mišković, "Probing the Plasmon-Phonon Hybridization in Supported Graphene by Externally Moving Charged Particles," *Plasmonics,* vol. 10, pp. 1741-1749, 2015/12/01 2015.
[87] V. W. Brar, M. S. Jang, M. Sherrott, S. Kim, J. J. Lopez, L. B. Kim, *et al.*, "Hybrid surface-phonon-plasmon polariton modes in graphene/monolayer h-BN heterostructures," *Nano letters,* vol. 14, pp. 3876-3880, 2014.
[88] C. R. Dean, A. F. Young, I. Meric, C. Lee, L. Wang, S. Sorgenfrei, *et al.*, "Boron nitride substrates for high-quality graphene electronics," *Nature nanotechnology,* vol. 5, pp. 722-726, 2010.
[89] H. Yan, T. Low, F. Guinea, F. Xia, and P. Avouris, "Tunable phonon-induced transparency in bilayer graphene nanoribbons," *Nano letters,* vol. 14, pp. 4581-4586, 2014.
[90] X. Lin, Y. Yang, N. Rivera, J. J. López, Y. Shen, I. Kaminer, *et al.*, "All-angle negative refraction of highly squeezed plasmon and phonon polaritons in graphene–boron nitride heterostructures," *Proceedings of the National Academy of Sciences,* vol. 114, pp. 6717-6721, 2017.
[91] Y. Qu, N. Chen, H. Teng, H. Hu, J. Sun, R. Yu, *et al.*, "Tunable Planar Focusing Based on Hyperbolic Phonon Polaritons in α-MoO3," *Advanced Materials,* p. 2105590, 2022.
[92] H. Hu, N. Chen, H. Teng, R. Yu, Y. Qu, J. Sun, *et al.*, "Doping-driven topological polaritons in graphene/α-MoO3 heterostructures," *Nature Nanotechnology,* vol. 17, pp. 940-946, 2022/09/01 2022.
[93] R. Fandan, J. Pedrós, J. Schiefele, A. Boscá, J. Martínez, and F. Calle, "Acoustically-driven surface and hyperbolic plasmon-phonon polaritons in graphene/h-BN heterostructures on piezoelectric substrates," *Journal of Physics D: Applied Physics,* vol. 51, p. 204004, 2018.
[94] H. Hajian, A. Ghobadi, S. A. Dereshgi, B. Butun, and E. Ozbay, "Hybrid plasmon–phonon polariton bands in graphene–hexagonal boron nitride metamaterials," *JOSA B,* vol. 34, pp. D29-D35, 2017.
[95] H. Song, S. Zhou, Y. Song, X. Wang, and S. Fu, "Tunable propagation of surface plasmon-phonon polaritons in graphene-hBN metamaterials," *Optics & Laser Technology,* vol. 142, p. 107232, 2021.
[96] L. Falkovsky and A. Varlamov, "Space-time dispersion of graphene conductivity," *The European Physical Journal B,* vol. 56, pp. 281-284, 2007.
[97] B.-C. Liu, L. Yu, and Z.-X. Lu, "The surface plasmon polariton dispersion relations in a nonlinear-metal-nonlinear dielectric structure of arbitrary nonlinearity," *Chinese Physics B,* vol. 20, p. 037302, 2011.
[98] F. W. Olver, D. W. Lozier, R. F. Boisvert, and C. W. Clark, *NIST handbook of mathematical functions hardback and CD-ROM*: Cambridge university press, 2010.
[99] A. G. Gurevich and G. A. Melkov, *Magnetization oscillations and waves*: CRC press, 1996.
[100] C. T. Kelley, *Solving nonlinear equations with Newton's method*: SIAM, 2003.
[101] D. M. Gay and R. B. Schnabel, "Solving systems of nonlinear equations by Broyden's method with projected updates," in *Nonlinear Programming 3*, ed: Elsevier, 1978, pp. 245-281.
[102] G. A. Sod, "A survey of several finite difference methods for systems of nonlinear hyperbolic conservation laws," *Journal of computational physics,* vol. 27, pp. 1-31, 1978.
[103] P. Berini, "Figures of merit for surface plasmon waveguides," *Optics Express,* vol. 14, pp. 13030-13042, 2006.
[104] H. Hu, R. Yu, H. Teng, D. Hu, N. Chen, Y. Qu, *et al.*, "Active control of micrometer plasmon propagation in suspended graphene," *Nature communications,* vol. 13, pp. 1-9, 2022.

[105] A. Woessner, M. B. Lundeberg, Y. Gao, A. Principi, P. Alonso-González, M. Carrega, *et al.*, "Highly confined low-loss plasmons in graphene–boron nitride heterostructures," *Nature materials,* vol. 14, pp. 421-425, 2015.

[106] S. Ye, Z. Wang, C. Sun, C. Dong, B. Wei, B. Wu, *et al.*, "Plasmon-phonon-polariton modes and field enhancement in graphene-coated hexagon boron nitride nanowire pairs," *Optics Express,* vol. 26, pp. 23854-23867, 2018.

[107] X. Yang, F. Zhai, H. Hu, D. Hu, R. Liu, S. Zhang, *et al.*, "Far-field spectroscopy and near-field optical imaging of coupled Plasmon–phonon polaritons in 2D van der Waals heterostructures," *Advanced materials,* vol. 28, pp. 2931-2938, 2016.